\documentclass[checkin,showpacs,aps,prl]{revtex4}
\usepackage{threeparttable}
\usepackage{amsmath}
\usepackage{multirow}
\usepackage{booktabs}
\usepackage{graphicx,color}
\usepackage{graphicx}
\usepackage{subfigure}
\usepackage{amsmath}

\newcommand{\ba}{\begin{eqnarray}}
\newcommand{\ea}{\end{eqnarray}}

\newcommand{\be}{\begin{equation}}
\newcommand{\ee}{\end{equation}}

\newcommand{\et}{{\it et al. }}
\newcommand{\ete}{{\it et al.}}

\definecolor{pink}{rgb}{1,0.18,1.0} 
\def\prl{{ Phys. Rev. Lett. }}

\def\apl{{ Appl. Phys. Lett. }}
\def\prb{{ Phys. Rev. B }}

\def\cpl{{ Chem.\ Phys.\ Lett.\ }}

\def\jpcm{{J. Phys.: Condens. Matter }}
\def\aip{{Aip Advances }}

\def\pnas{{Proc. Natl. Acad. Sci. USA }}
\def\nm{{Nature Mat. }}
\begin{document}

\title{Manipulating femtosecond magnetism through pressure:\\
 First-principles calculation}

\author{M. S. Si} \affiliation{Key Laboratory for Magnetism and
Magnetic Materials of the Ministry of Education, Lanzhou University, 
Lanzhou 730000, China \& \\ Department of Physics, Indiana State
University, Terre Haute, Indiana 47809, USA}

\author{J. Y. Li}
\affiliation{Key Laboratory for Magnetism and Magnetic Materials of
  the Ministry of Education, Lanzhou University, Lanzhou 730000,
  China}

\author{D. S. Xue$^*$}
\affiliation{Key Laboratory for Magnetism and Magnetic Materials of
  the Ministry of Education, Lanzhou University, Lanzhou 730000,
  China}

\author{G. P. Zhang$^{\star}$}
\affiliation{Department of Physics, Indiana State University, Terre
  Haute, Indiana 47809, USA}


\newcommand{\OSGe}{$\cal S\bigotimes D$}
\newcommand{\OSG}{$\cal S\bigotimes D$ }
\newcommand{\osg}{{\cal S\bigotimes D} }
\newcommand{\sd}{$\cal SD$ }

\date{\today}

\begin{abstract}
   { Inspired by a recent pressure experiment in fcc Ni, we propose a simple method
     to use pressure to investigate the laser-induced femtosecond magnetism.
     Since the pressure effect on the electronic and magnetic properties
     can be well controlled experimentally, this leaves little room
     for ambiguity when compared with theory. Here we report our theoretical
     pressure results in fcc Ni: Pressure first suppresses the spin moment reduction, and 
     then completely diminishes it; further increase in pressure to 40 GPa
   induces a demagnetization-to-magnetization transition.  To reveal
   its microscopic origin, we slide through the L-U line in the Brillouin
   zone and find two essential transitions are responsible for this
   change, where the pressure lowers two valence bands, resulting in
   an off-resonant excitation and thus a smaller spin moment
   reduction.  In the spin-richest L-W-W$'$ plane, two spin contours
   are formed; as pressure increases, the contour size retrieves
   and its intensity is reduced to zero eventually, fully consistent
   with the spin-dipole factor prediction.  These striking features
   are detectable in time- and spin-resolved photoemission
   experiments.}
\end{abstract}
 \maketitle

\section{I. Introduction}

Laser-induced femtosecond (de)magnetism dynamics opens a new frontier
in magnetism, femtomagnetism \cite{beaurepaire}. Interest in femtomagnetism has grown
substantially \cite{aes,comin,vap,kir,koo,kam,ourreview,sijpcm} due to its fundamental importance and
practical applications \cite{pickel,stanciu,bigotnature,mathias}.
However, its underlying mechanism is quite different from those driven
by magnetic or thermal fields, because the electric field of a laser
directly interacts with charges of a sample, but only indirectly with
the spin.  A systematic and well controlled approach is a must.  It
was a recent pressure experiment that caught our attention. Torchia
{\ete} \cite{torchio} carried out a pressure-dependent x-ray magnetic
circular dichroism (XMCD) measurement in fcc Ni. They showed that the
ferro- versus para-magnetic phase transition does not occur at 160 GPa
as predicted theoretically, and suggested the disappearance of
ferromagnetic phase at 400 GPa \cite{pressnote}. Although this experiment has nothing
to do with femtomagnetism, it suggests a possibility to manipulate the
ultrafast magnetization through pressure.  First, the experiment
showed that Ni retains its pristine fcc structure under pressure. This
presents a clean case, free of the symmetry-broken effects, where a
direct comparison between experiment and theory is possible.  Second,
XMCD has already been developed into a time-resolved technique on the
femtosecond time scale \cite{stamm}.  Theoretical predictions can be
tested experimentally.  Up to now, neither experimental nor
theoretical investigation has ever been carried out in laser-induced
femtomagnetism in any materials under pressure. We believe that a
theoretical study is timely and appropriate.

In this work, we aim to demonstrate that the pressure is a promising
tool to investigate the laser-induced ultrafast spin dynamics. Our
first-principles calculation shows that pressure introduces a clean
and dramatic change in (de)magnetization.  In fcc Ni, it first
decreases the spin moment reduction, and then erases it entirely. If
the pressure is increased further to 40 GPa, it induces a
demagnetization-magnetization transition, a finding that has never
been reported in metals.  To reveal its microscopic origin, we examine
the L-U line in the Brillouin zone (BZ) and find that two transitions
are responsible for these dramatic changes. These two transitions are
from two valence bands, each of which is 2 eV below the Fermi level,
to a conduction band, which is just above the Fermi level. When the
pressure increases, the valence bands are downshifted with respect to
the conduction band, leading to an off-resonant transition and a
smaller spin moment change. The largest spin moment change resides in
the L-W-W$'$ plane, where the spin moment contour starts with two strong
concentric arcs at 0 GPa; as pressure increases, it loses its
intensity, in agreement with the spin-dipole factor prediction.  We
emphasize that all the transitions are gradual, without a sudden
crystal symmetry changes, thus it is particularly suitable for the
experimental investigation.  This paper is organized as
follows. Section II is devoted to the theoretical formalism. In
Sec. III, we discuss the obtained results. Finally, we conclude this
paper in Sec. IV.

\section{II. Theoretical formalism}

Pressure is a powerful tool to manipulate material properties
\cite{ols,loa}.  {\color{black}
It can induce the superconductivity in cerium
\cite{wittig}, bcc barium \cite{pro}, $\rm UPt_3$ \cite{joynt,Chan},
antiferromagnetism in invar \cite{invar}, and magnetic phase changes
in PrSb \cite{vet}, SrFeO$_3$ and BaFeO$_3$ \cite{li,boh}, and
Nd$_{0.53}$Sr$_{0.47}$MnO$_3$ \cite{bal}.}  Pressure has been used to
control ultrafast dynamics before. Employing a 60 fs/620 nm laser
pulse, Lienau {\ete} \cite{lienau} found that in the compressed
supercritical argon, the femtosecond transients of iodine undergo a
big change with argon pressures.  Very recently Liu \et \cite{liu}
showed that the lifetime of the S-1 state in beta-carotene increases
with pressure. Kasami \et \cite{kasami} found that in bismuth the high
pressure suppresses the amplitude of coherent phonons and changes
their frequencies. They assigned this to the disappearance of the
band-structure overlapping at L- and T-points.  In metal nanoparticles
\cite{christofilos}, the vibration frequency increases with pressure.
It is certainly of great interest to see whether the pressure could
introduce similar dramatic effects in femtomagnetism. It is well known
that high pressure generated by the so-called 
diamond anvil cell technique is very challenge in experiments. Here we
instead suggest an alternative method, that is epitaxy-induced lattice
change, to mimic this effect.  If there is a misfit between a
substrate and a growing epilayer, the first atomic layers which are
deposited will be  strained to match the substrate \cite{people}. As a result, the
corresponding lattice constant will be enlarged or reduced which
depends on the substrate used. Such a substrate-induced lattice change
can be regarded as an equivalent means to mimic the effect of pressure. Importantly, the
measurement of femtosecond laser induced magnetization can be carried
out on those epitaxial thin films. Figure
\ref{fig1}(a) shows a schematic of our proposed pressure experiment,
the band dispersion and spin moment change along the L-U direction
(see below for more).

Our theoretical calculation is based on the density functional theory
and employs the full potential augmented planewave method as
implemented in the WIEN2k code \cite{wien}.  To accurately take into
account the diffusive nature of the Kohn-Sham excited states, we
choose a large product of the Muffin-tin radius and the planewave
cutoff of 9.5. Additional linearization energies are added to
represent those high lying states where the laser accesses. Spin-orbit
coupling is included. Spin matrices are constructed by ourselves
\cite{prb09}. Both our spin moment and band structure agree with the
prior experimental and theoretical results \cite{vw}.
There are several similarities and differences
between the work by Vernes and Weinberger and ours. They include both
the current-current correlation, which is similar to our single pulse
excitation, and three-current correlation, which gives the additional
dissipation. They linearized their equation of motion  with
respect to the probe pulse, while we directly solve the Liouville
equation numerically.
 Whereas the electron correlation
effect is important to the spin 
  moment change, as a first step, in this paper we focus on the
  pressure effect on magnetization within the single particle
  picture. In addition, we also ignore the lattice effect as several
  theoretical studies show its effect is small \cite{lefkidis, carva,
    kraub} at least in the early stage at laser excitation.

\section{III. Results and Discussion}

\begin{figure}
\includegraphics[width=16cm]{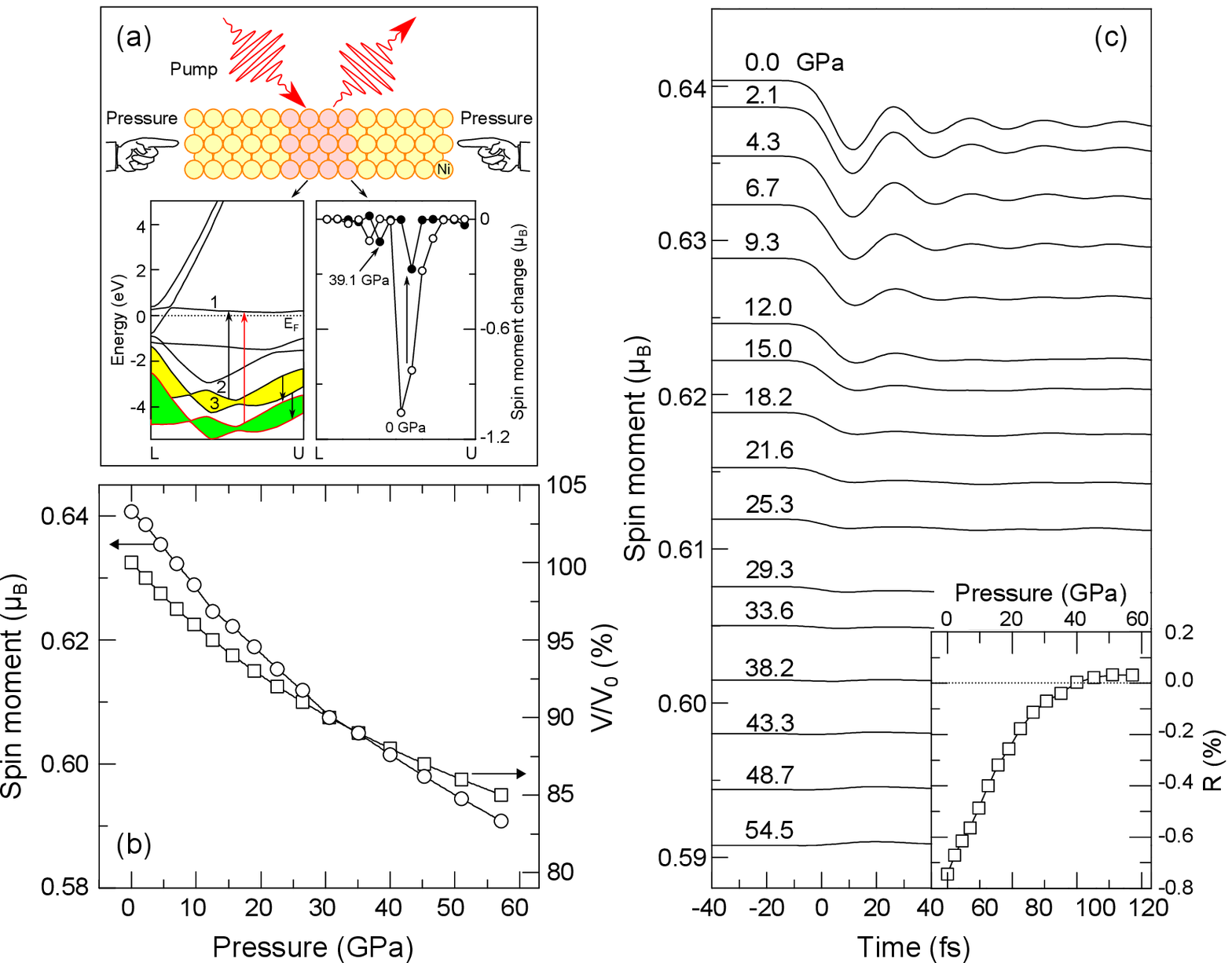}
\caption{(color online). (a) Schematic of femtosecond magnetism under
  pressure. Bottom left: Band structure along the L-U line, where two
  transitions 2$\rightarrow$1 and 3$\rightarrow$1 are responsible for
  the spin moment change. As pressure increases from 0 to 39.1 GPa,
  the valence bands 2 and 3 are downshifted from the yellow to green
  bands.  Bottom right: Spin moment change along the L-U line obtained
  at 0 (empty circles) and 39.1 GPa (filled circles), respectively.
  (b) Volume change with pressure (squares), data fitted to the
  Birch-Murnaghan equation of state, and the spin moment change with
  pressure (circles).  (c) Time evolution of the spin moment at
  pressures 0.0, 2.1, 4.3, 6.7, 9.3, 12.0, 15.0, 18.2, 21.6, 25.3,
  29.3, 33.6, 38.2, 43.3, 48.7, and 54.5 GPa from top to bottom,
  respectively. Inset: Maximum magnetization change rate $\cal R$ as a
  function of pressure. The dotted line refers the zero $\cal R$. }
\label{fig1}

\end{figure}

Before we carry out dynamical simulations, we first verify the
pressure dependence of the volume (see the boxes in
Fig. \ref{fig1}(b)). Our interest is in the spin moment change with
pressure.  Figure \ref{fig1}(b) (the circles) shows that as pressure
increases, the magnetic moment decreases monotonically from 0.64
$\mu_{\rm B}$ at 0 GPa to 0.59 $\mu_{\rm B}$ at 54.5 GPa, in good
agreement with the results by Torchio \et \cite{torchio}.  Upon
ultrafast laser excitation, the spin moment undergoes a dramatic
change. Beaurepaire \et \cite{beaurepaire} showed that their
time-resolved magneto-optical Kerr signal in fcc Ni dropped
precipitously within 1 ps. Subsequent experiments soon verified their
findings. However, to simulate such a fast process is no easy task
\cite{naturephysics2009,prl00,prb09}, because it involves a real time
evolution with a huge number of $k$ points. To solve this problem, we
employ the Liouville equation for the density matrix $\rho$ at each
${\bf k}$ \cite{naturephysics2009}, \be i\hbar(\partial\rho_{\bf
  k}/{\partial t})=[H_0+H_I,\rho_{\bf k}],\ee where $H_0$ is the
system Hamiltonian, and $H_I$ is the interaction between the laser and
the system, i.e., $H_I=\sum_{k;i,j}\vec{D}_{ij} \cdot \vec{E}(t)
c_{k,i}^\dag c_{k,j}$.  Here $\vec{D}$ is the dipole operator, and
$c_{k,i}^\dag$ ($c_{k,i}$) is the electron creation (annihilation)
operator for band $i$ at $k$. The laser field is $\vec{E}(t)=\hat{e}
A_0 \exp{[-t^2/\tau^2]} \cos(\omega t)$, where $\hat{e}$ is the laser
polarization direction, $A_0$ is the field amplitude, $\omega$ is the
laser frequency, $t$ is the time and $\tau$ is the laser pulse
duration. This is a coupled equation involving transitions between all
the band states and is solved numerically using massively parallel
computers.  Once we obtain the density matrix, the time-dependent spin
moment can be computed by tracing over the product of the density and
spin matrices.

We choose the photon energy of 2 eV, the laser pulse duration of 12 fs
and the field amplitude of 0.05 V/{\AA}.  Figure 1(c) shows the spin
moment change as a function of time at sixteen pressures from 0 GPa
through 54.5 GPa.  A few striking features are noticeable.  First, the
amount of demagnetization is reduced with pressure.  For the
time-resolved magneto-optical Kerr experiment, this means that the
Kerr signal will drop with pressure.  For instance, at 0 GPa, the spin
change is -0.0048 $\mu_{\rm B}$, while at 23 GPa, it is -0.0011
$\mu_{\rm B}$.  Second, as pressure increases, the strong oscillation
in the spin moment is greatly reduced.  The period of this spin
oscillation is determined by the spin-orbit coupling \cite{jap08}.
Once the pressure is above 18 GPa, the oscillation nearly
diminishes. This theoretical result can be tested
experimentally. Thirdly, and surprisingly, when the pressure increases
further, there is a demagnetization-to-magnetization transition at 40
GPa, where the system is now magnetized, instead demagnetized. Such a
magnetization enhancement has been observed in ferromagnetic
semiconductors \cite{chemlaprl0405,fliu} and metallic multilayers
\cite{rud}, but has never been reported in ferromagnetic metals.  To
quantify these changes, we introduce the percentage change as \be
{\cal R} = \frac{M_{\rm ex}-M(-\infty)}{M(-\infty)}, \ee where $M_{\rm
  ex}$ is the spin moment extreme (maximum or minimum) and
$M(-\infty)$ is the initial magnetic moment prior to the femtosecond
laser pulse. A positive or negative $ {\cal R}$ corresponds to the
magnetization or demagnetization, respectively.  The inset of
Fig. 1(c) shows that as pressure increases, $\cal R$ drops from
$-$0.74\% to zero, and changes its sign around 40 GPa, before it
finally saturates at 0.03\% \cite{note}. We note that at ambient pressure and for the same
realistic electric field, the amount of demagnetization is much
smaller than that observed in experiments. This might be due to the
neglect of electron-electron interaction and the system difference,
where our system is bulk fcc Ni, but most experimental systems are
thin films. On the picosecond or longer timescale, this may become
necessary to include the electron-phonon interaction.

\begin{figure}
\includegraphics[width=16cm]{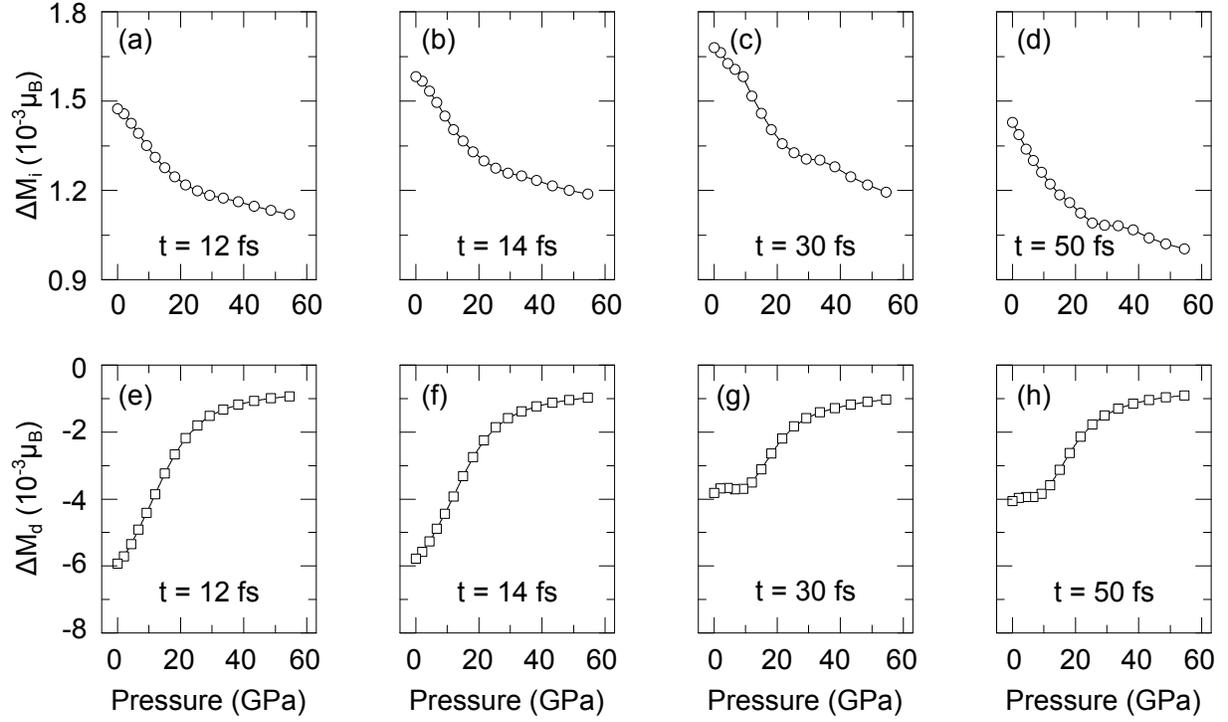}
\caption{(color online). Effect of pressure on the spin moment change
  at four selected times, 12, 14, 30, and 50 fs.  Magnetization change
  is sorted into two groups: (a)-(d) represent the spin moment
  increase and (e)-(f) represent the spin moment decrease.  }
\label{fig2}

\end{figure}

To understand how the pressure induces such a dramatic change, we
first separate all the $k$ points into two groups: One with the spin
moment increased ($\Delta M_i$) and the other with the spin moment
decreased ($\Delta M_d$).  Then we plot $\Delta M_i$ and $\Delta M_d$
as a function of pressure at four selected times, 12, 14, 30 and 50
fs, where the extremes appear.  Since the net magnetization change is
simply a sum of these two, it is useful to examine how each changes
with pressure. Figures \ref{fig2}(a)-(d) represent the magnetization
increase, while Figs. \ref{fig2}(e)-(h) the magnetization decrease.  A
common feature emerges: Independent of the times selected, both
$|\Delta M_i|$ and $|\Delta M_d|$ decrease with pressure. However, the
amount of the reduction is different. The net change in $|\Delta M_i|$
is about $0.45\times 10^{-3}\mu_{\rm B}$, whereas the change in
$|\Delta M_d|$ is about $6\times 10^{-3}\mu_{\rm B}$, more than ten
times larger than $|\Delta M_i|$. That is why at pressures below 40
GPa, the entire spin moment change follows the trend set by $|\Delta
M_d|$. Once above 40 GPa, $|\Delta M_d|$ is reduced so much that it
can not compete with $|\Delta M_i|$; consequently, $|\Delta M_i|$
completely dominates the spin moment change. If $|\Delta M_i|=|\Delta
M_d|$, we have zero spin moment change. On the other hand, if $|\Delta
M_i|>|\Delta M_d|$, the system becomes magnetized, instead
demagnetized. This is the origin of the pressure-dependent spin moment
change. To pin down its microscopic picture, we focus on a few
critical lines in the Brillouin zone.

\newcommand{\wsd}{$\tilde{\cal SD}$\ }

\begin{figure}
\includegraphics[width=15cm]{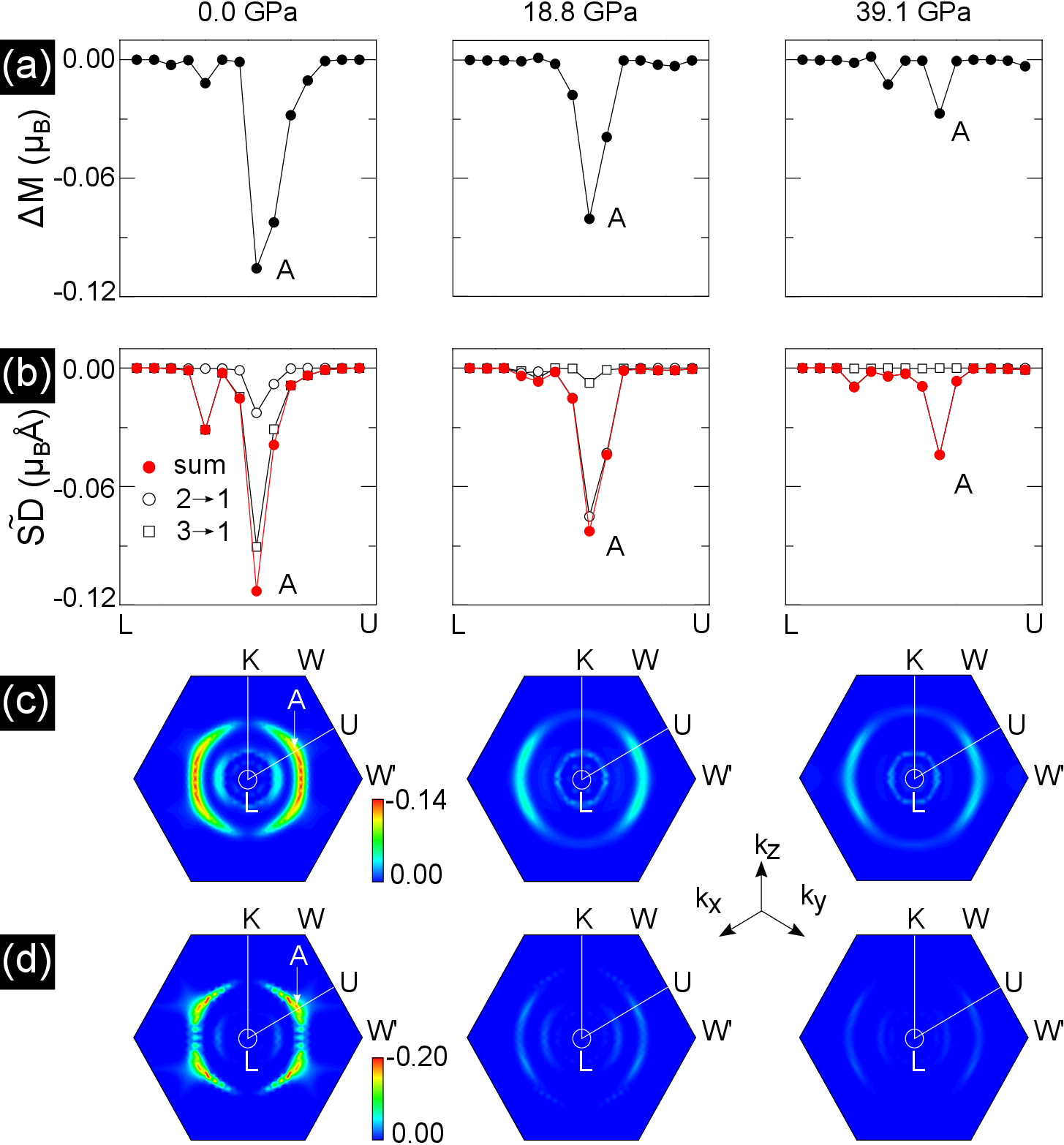}
\caption{(color online). Crystal-momentum resolved spin moment change
  and \wsd factor change along the L-U line and within the L-W-W$'$
  plane at three pressures, 0, 18.8 and 39.1 GPa. (a) Dispersed spin
  moment change with pressure along the L-U line. The maximum change
  is {\color{black} near the} point $A$
  [$\frac{2\pi}{a}(80,38,-38)/104$]. (b) Energy-weighted \wsd factor
  diagram.  The unfilled circles refer to the contribution from the
  transition from state 2 to 1 (see Fig. \ref{fig1}(a)), the boxes
  refer to the contribution from state 3 to state 1 (see
  Fig. \ref{fig1}(a)), and the filled circles refer to the sum of the
  above two contributions. (c) Contour of the \wsd factor change with
  pressure in the L-W-W$'$ plane.  (d) Contour of the spin moment change
  in the L-W-W$'$ plane.  }
\label{fig3}

\end{figure}

One such line is the L-U line, where a strong spin change is
observed. We crystal-momentum-resolve the spin moment change from the
L to U point at three pressures from 0 through 39.1 GPa (see the three
figures in Fig. \ref{fig3}(a)). Increasing pressure narrows the
profile of the spin moment change and reduces the spin moment change
at all the $k$ points. This spectacular change must be connected with
the band structure change, which is shown in Fig. \ref{fig1}(a) with
the Fermi energy at 0 eV. Based on the photon energy of 2 eV, we
identify two transitions directly responsible for the spin moment
change: One is from valence band 2 to conduction band 1 and the other
is from valence band 3 to conduction band 1 (see the band dispersion
in Fig. \ref{fig1}(a)). As shown before \cite{siaip}, what matters to
the spin moment change is the transition matrix elements and the
intrinsic spin differences between transition states.  To capture
their composite effect, we resort to the \sd factor \cite{siaip}, \be
({\cal SD})_{ij}=\Delta S_{ij} D_{ij} \ee where $i$ and $j$ are state
indices, $\Delta S$ is the spin difference between these two states
and $D$ is the transition matrix element between them.  We find that the \sd
factor does change with pressure, but the change is too small
to explain the change in Fig. \ref{fig3}(a) (not shown). The missing
piece of the puzzle is the transition energy change.  As pressure
increases, both valence bands 2 and 3 are downshifted from the yellow
to green bands (see the bottom left inset of Fig. \ref{fig1}(a)). This
shift is detrimental to the spin change since now the transition
energy falls beyond the photon energy of 2 eV and the transition
becomes off-resonant.  To include the transition energy effect, we
multiply the \sd factor by the Gaussian function \be
W(E)=\exp[-(E-E_{\rm photon})^2/\sigma^2] \ee to get a weighted
$\tilde{\cal SD}$ factor.  Here $E$ is the pressure-dependent
transition energy, $E_{\rm photon}$ is the photon energy and
$\sigma=0.125$ eV is a small broadening.  Our transition-energy
weighted $\tilde{\cal SD}$ factor diagram is shown in
Fig. \ref{fig3}(b). Now, if we compare Figs. \ref{fig3}(a) and
\ref{fig3}(b), we see that the \wsd factor diagram now agrees with the
spin moment change and both peak {\color{black} near} a point $A$
[$\frac{2\pi}{a}(80,38,-38)/104$] where $a$ is the lattice constant of
fcc Ni. As pressure increases, this peak shifts slightly toward a
larger momentum. This demonstrates that the pressure-induced spin
moment change is a joint effect of the spin moment difference,
transition matrix element and laser excitation energy window.


We want to know whether the above explanation also holds true for the
entire Brillouin zone.  We choose a representative plane, the L-W-W$'$
plane, defined by five special $k$ points, L, K, W, U and W$'$. This
plane has the strongest demagnetization.  We crystal-momentum resolve
the \wsd in the two-dimensional plane.  Similar to the photoemission
intensity map, Fig. \ref{fig3}(c) is an amplitude contour of \wsd
change, with the scales shown in the lower right corner. Our results
are insightful. First, the \wsd is not uniform across the plane. Those
points with a large \wsd are clustered in two sectors, defined by two
L-U lines, and they form two concentric arcs. Points with the
strongest changes are in the middle, facing the U, W$'$ and U points.
Point $A$ in Fig. \ref{fig3}(b) is now just one of many hot spin
points in Fig. \ref{fig3}(c). Second, as pressure increases two arcs
evolve dramatically.  Energetically, two arcs are 0.70 and 1.94 eV
from the Fermi surface (see the white circle around the L point). This
separation is due to the finite photon energy imposed on the \wsd
factor. An increase in pressure reduces the \wsd factor. The
amplitudes of both the inner and outer arcs drop strongly. The inner
arc retrieves toward the Fermi surface. Further increase in pressure
greatly reduces the amplitudes. We see the same pattern occurring in
the spin moment change with pressure (see
Fig. \ref{fig3}(d)). However, there are some minor differences.  The
\wsd factor has much finer structures in it, while the spin change
contour shows much smoother variations in the momentum space. Also, we
see that the spin change is concentrated at a point or its equivalence
along the L-U line. These differences are due to the laser excitation,
where the pulse smears out the fine details. Except these small
differences, our \wsd factor captures the main picture of the
dynamical spin moment changes.  We note in passing that our result
remains qualitatively same for a 24 fs pulse. Therefore, our
pressure-induced spin change greatly broadens our view of
femtomagnetism. Since the state-of-the-art time-, momentum- and
spin-resolved photoemission techniques \cite{pickel,stamm} are already
available, we expect a direct experimental test of our prediction.
In high-temperature superconductors, ultrafast lasers have long been used to
investigate the Cooper pairing \cite{smallwood}. Pressure can be directly applied to those
materials, in a similar fashion as discussed here.

\section{IV. Conclusion}

We have proposed a new method to investigate the spin
dynamics in ferromagnets, and demonstrated theoretically that the
pressure has a dramatic effect on the laser-induced femtosecond
(de)magnetization process. This technique is simple and
straightforward, an ideal test case for a detailed experimental and
theoretical comparison. The pressure greatly decreases the spin moment
reduction and then diminishes it as the pressure increases
further. The reason is that the magnetization and demagnetization
pockets depend on the pressure differently. {\color{black}The
  demagnetization is very susceptible to the pressure, but not for the
  magnetization.} This creates a competition between them. When the
pressure is above 40 GPa, the magnetization pocket completely
dominates; as a result, a magnetization enhancement is observed and
the demagnetization-to-magnetization transition occurs. The
microscopic origin is revealed by examining the band structure change
with the pressure. We find that two transitions from two different
valence bands to the same conduction are responsible for the dramatic
change along the L-U line.  This explains the evolution of the most
intense spin change pockets in the L-W-W$'$ plane. Future experiments
can directly test our predictions.

\section{Acknowledgments}

This work was supported by the National Basic Research Program of
China under No. 2012CB933101 and the U. S. Department of Energy under
Contract No. DE-FG02-06ER46304 (MSS and GPZ). This work was also
supported by the National Science Foundation of China (NSFC) under
Nos. 10804038, 11034004, and 50925103 and the Fundamental Research 
Funds for the Central Universities (No. 2022013zrct01).
We acknowledge part of the work as done on Indiana
State University's high-performance computers.  This research used
resources of the National Energy Research Scientific Computing Center
at Lawrence Berkeley National Laboratory, which is supported by the
Office of Science of the U.S. Department of Energy under Contract
No. DE-AC02-05CH11231. Our calculation also used resources of the
Argonne Leadership Computing Facility at Argonne National Laboratory,
which is supported by the Office of Science of the U. S. Department of
Energy under Contract No. DPE-AC02-06CH11357. MSS thanks
the State Scholarship Fund by the China Scholarship Council for
financially supporting his visit at the Indiana State University.

$^*$xueds@lzu.edu.cn

$^\star$gpzhang@indstate.edu

\clearpage

\clearpage

\clearpage


\begin{thebibliography}{99}
\bibitem{pressnote} Though one cannot rule out a possible crystallographic
phase transition or a transition to a paramagnetic state at ambient temperature
and at lower pressure \cite{torchio} in experiments, fcc Ni phase
with ferromagnetic order below $\sim$400 Gpa is predicted theoretically, which
is also consistent with our simulated results based on the density functional 
theory.
\end{thebibliography}
\end{document}